\documentclass[twocolumn,10pt]{asme2ej}

\usepackage{epsfig} 
\usepackage{lipsum}

\usepackage{epsfig}
\usepackage{amsmath}

\usepackage{amsthm}
\usepackage{amsfonts}
\usepackage{amssymb}
\usepackage{mathdots}
\usepackage{mathdots}
\usepackage[utf8]{inputenc}
\usepackage[titletoc,title]{appendix}
\usepackage{titletoc}

\usepackage{enumerate}
\usepackage{centernot}
\usepackage[mathscr]{euscript}
\usepackage{url}
\usepackage{hyphenat}
\usepackage{bbm}
\usepackage{bm}
\usepackage{verbatim}
\usepackage{accents}
\usepackage{graphicx}
\usepackage{subcaption}
\usepackage{listings}
\usepackage{xcolor}
\usepackage{grffile}
\usepackage{rotating}
\usepackage{tikz}

\usepackage{grffile}
\usepackage{algorithm,algorithmicx}
\usepackage{algpseudocode}
\algnewcommand\algorithmicinput{\textbf{Input: }}
\algnewcommand\algorithmicoutput{\textbf{Output: }}
\newcommand\E[1]{{\mathbb{E}\left[#1\right]}}

\usepackage{afterpage}
\usepackage{epstopdf}
\usepackage{widetext}
\usepackage{cuted}
\usepackage{multicol}
\everymath{\displaystyle}
\DeclareMathOperator*{\argmin}{argmin}
\DeclareMathOperator*{\argmax}{argmax}

\newcommand\blue[1]{\textcolor{blue}{\text{#1}}}
\newcommand\serif[1]{\textsf{#1}}

\usepackage{lineno,hyperref}
\modulolinenumbers[5]

\theoremstyle{definition}
\newtheorem{theorem}{Theorem} 
\newtheorem{definition}[theorem]{Definition}

\theoremstyle{remark}

\usepackage{stackengine,graphicx}
\stackMath
\newlength\matfield
\newlength\tmplength
\def\matscale{1.}
\newcommand\dimbox[3]{%
  \setlength\matfield{\matscale\baselineskip}%
  \setbox0=\hbox{\vphantom{X}\smash{#3}}%
  \setlength{\tmplength}{#1\matfield-\ht0-\dp0}%
  \fboxrule=1pt\fboxsep=-\fboxrule\relax%
  \fbox{\makebox[#2\matfield]{\addstackgap[.5\tmplength]{\box0}}}}
\newcommand\raiserows[2]{%
   \setlength\matfield{\matscale\baselineskip}%
   \raisebox{#1\matfield}{#2}}
\newcommand\matbox[5]{
  \raisebox{2.3ex}{\rotatebox[origin=center]{90}{\scalebox{.45}{#2}}}\,%
  \stackon{\dimbox{#1}{#3}{#5}}{\scalebox{.45}{#4}}}
\newcommand\titlebox[3][30ex]{\stackanchor{#2}{\scalebox{.45}{\parbox{#1}{\centering #3}}}}

\title{Scalable$^3$-BO: Big Data meets HPC - A scalable asynchronous parallel high-dimensional Bayesian optimization framework on supercomputers}

\author{Anh Tran\thanks{Corresponding author: Anh Tran (anhtran@sandia.gov)}
    \affiliation{
    Optimization and Uncertainty Quantification \\ 
    Sandia National Laboratories \\
    Albuquerque, NM 87123 \\
    Email: anhtran@sandia.gov
    }
}

\begin{document}

\maketitle  
\makeatletter
\let\ps@oldempty\ps@empty 
\renewcommand\ps@empty\ps@plain
\makeatother


\begin{abstract}
Bayesian optimization (BO) is a flexible and powerful framework that is suitable for computationally expensive simulation-based applications and guarantees statistical convergence to the global optimum. While remaining as one of the most popular optimization methods, its capability is hindered by the size of data, the dimensionality of the considered problem, and the nature of sequential optimization. These scalability issues are intertwined with each other and must be tackled simultaneously. In this work, we propose the Scalable$^3$-BO framework, which employs sparse GP as the underlying surrogate model to scope with Big Data and is equipped with a random embedding to efficiently optimize high-dimensional problems with low effective dimensionality. The Scalable$^3$-BO framework is further leveraged with asynchronous parallelization feature, which fully exploits the computational resource on HPC within a computational budget. As a result, the proposed Scalable$^3$-BO framework is scalable in three independent perspectives: with respect to data size, dimensionality, and computational resource on HPC. The goal of this work is to push the frontiers of BO beyond its well-known scalability issues and minimize the wall-clock waiting time for optimizing high-dimensional computationally expensive applications. We demonstrate the capability of Scalable$^3$-BO with 1 million data points, 10,000-dimensional problems, with 20 concurrent workers in an HPC environment.
\end{abstract}

\section{Introduction}

Bayesian optimization (BO) is an efficient optimization method for computationally expensive and complex real-world engineering applications. It is constructed up on an underlying Gaussian process (GP), which is adaptively refined as the optimization process advances. 
Although BO has been used extensively in the literature, there are multiple drawbacks that limits the capability of BO methods, mainly because of different scalability issues. First, it is well-known that training GP costs $\mathcal{O}(n^3)$ flops to compute the inverse of the covariance matrix, which limits the number of observations to 10$^2$-10$^4$ depending on the specific implementation. This is usually referred to as the data scalability issue. 
Second, GP does not typically perform well in high-dimensional problems because training a GP is, again, an optimization problem on high-dimensional space to search for the best hyper-parameters. 
Third, the classical BO algorithm is sequential; that is, it only queries one simulation or one functional evaluation at a time, which in turn creates a computational bottleneck for computationally expensive problems.

In this paper, we propose a robust and scalable approach to solve a high-dimensional Bayesian optimization for computationally expensive applications by exploiting the computational resources on supercomputers. 
The proposed BO algorithm, called Scalable$^3$-BO, is scalable in three distinct directions. First, it is scalable with respect to the dataset. Second, it is scalable with respect to the dimensionality of the problem. Third, it is scalable with respect to the computational resource. 
This problem is important in so many aspects, as it leverages the capability of the classical BO methods and expands the applicability to a broader domain, particularly for computationally expensive high-dimensional modeling and simulation applications. 
While the idea of scalability seems to be disjoint in three different directions, they are closely related to each other. 
For example, optimizing in HPC will presumably generate a lot of data, so parallelization in HPC is tied to data scalability issue. 
High-dimensional problems also requires exponentially more data points to learn, typically, so it is also related to data scalability issue. 
Finally, it is difficult to solve a high-dimensional optimization problem without the usage of HPC, especially for computationally expensive applications in the era of computers. These intertwined issues therefore requires a comprehensive approach that deals with all scalability problems simultaneously. 


Local GP approaches is arguably the most popular choice for addressing data scalability due to its ease in implementation. 
Bostanabad et al. \cite{bostanabad2019globally} proposed a globally approximate local GP (GAGP) and demonstrated up to $\sim$90k data points. 
Zhang et al. \cite{zhang2020remarks} proposed a locally weighted scheme, similarly to van Stein et al. \cite{van2015optimally,van2016fuzzy}, where the weights are derived by minimizing the weighted posterior variance, and demonstrate to 100k data points. 
Tran et al. \cite{tran2018gpdft,tran2018efficient} proposed a local GP approach with Wasserstein distance to model the potential energy surface and mixed-integer BO problems \cite{tran2019constrained}. 
Nguyen et al. \cite{nguyen2009model} proposed to tune the weights adaptively as a function of inputs and cluster centers and demonstrated up to $\sim$15k data points. 
Keyes et al. \cite{abdulah2017exageostat,akbudak2017tile,abdulah2018parallel,litvinenko2019likelihood} proposed and employed a tile algorithm with hierarchical low-rank Cholesky approximation and demonstrated up to $\sim$2M data points with heterogeneous computing, i.e. CPU+GPU. 
Recently, Chen et al. \cite{chen2020stochastic} proposed to use stochastic gradient descent to train GP and demonstrated with 2M data points. 
Eriksson et al. \cite{eriksson2019scalable} also used local surrogate model to scale the global surrogate model. 
Liu et al.~\cite{liu2020gaussian} provided a comprehensive review of scalable GPs for Big Data. 

High-dimensional GP is also an interesting topic, which has attracted many researchers. Due to the limited space, we only select a few notable works, but much more have been done in the literature. Bach \cite{bach2009high} proposed a hierarchical kernel learning to select a hull of interaction terms. Duvenaud et al. \cite{duvenaud2011additive,durrande2012additive} proposed additive kernels, which strikingly resembles high-dimensional model representation and ANOVA decomposition. Wang et al. \cite{wang2013bayesian,wang2016bayesian} proposed REMBO, which draws a normal random matrix and optimizes the high-dimensional function in an embedded low-dimensional subspace. Nayebi et al. \cite{nayebi2019framework} proposed HeSBO, which can be considered as an extension of REMBO but differs in the embedding function. Li et al. \cite{li2018high} coupled Dropout to randomly select active variables. Li et al. \cite{li2016high} proposed a restricted projection pursuit model that has a similar flavor to active subspaces. 
Parallelism in BO also offers another boost to increasing optimization efficiency on high-performance computing (HPC) platforms. Relevant to this work, Desautels et al. \cite{desautels2014parallelizing} proposed GP-BUCB and GP-AUCB framework that focuses on the acquisition function, whereas Contal et al. \cite{contal2013parallel} proposed a GP-UCB-PE. \cite{kandasamy2017asynchronous} to promote exploration.

In this work, we take the sparse GP approach instead of the local GP approach as an underlying GP due to its solid theoretical foundation. While the complexity is more expensive compared to the local GP approach ($\mathcal{O}(nm^2)$ versus $\mathcal{O}(m^3)$), it allows better approximation accuracy in high-dimensional problems. 
We combine several aspects of scalability into a unified framework that is suitable for a wide range of single-objective expensive applications. 
In this paper, we propose Scalable$^3$-BO algorithm, which 
\begin{itemize}
\item - is scalable with respect to the size of data by adopting the sparse GP,
\item - is equipped with a random embedding to solve high-dimensional problems with low-dimensional active subspace,
\item - is fully parallelized with an asynchronous parallelism for computationally expensive simulation-based problems on HPC platforms.
\end{itemize}
This high-dimensional research direction is motivated by the promising results of active subspace, where even high-fidelity and computationally expensive simulations are shown to have active subspaces and low intrinsic dimensionality \cite{constantine2015active}. However, instead of finding the active subspace and accurately approximating the high-dimensional function, in this paper, we narrow the scope of our interest to optimize a high-dimensional problem that has an unknown low-dimensional active subspace as efficiently as possible. 
We benchmarked and demonstrated the optimization results in 10,000D and 1M dataset, with a smart scheduler and tracker in HPC environments with 20 concurrent workers.

\section{Bayesian optimization: Methodology}

For the sake of clarity and consistency, the following symbols are used and annotated throughout the paper.

\begin{itemize}
\item $\mathbf{x} \in \mathcal{X} \subset \mathbb{R}^D$: inputs,
\item $\mathbf{z} \in \mathcal{Z} \subset \mathbb{R}^d$: random embedded inputs,
\item $\mathbf{X_u} \in \mathcal{X} \subset \mathbb{R}^D$: inducing inputs,
\item $\mathbf{Z_u} \in \mathcal{Z} \subset \mathbb{R}^d$: random embedded inducing inputs,
\item $\mathbf{u} \in \mathcal{R}$: inducing random embedded outputs,
\item $\mathbf{y} \in \mathbb{R}$: outputs,
\item $D$: dimensionality of $x$ (before embedding),
\item $d \ll D$: dimensionality of $z$ (after embedding),
\item $\mathbf{A} \in \mathbb{R}^{D \times d}$: normal random matrix.
\end{itemize}

\subsection{Classical Gaussian Process and Bayesian optimization}

Assume that $f$ is a function of $\mathbf{x}$, where $\mathbf{x} \in \mathcal{X}$ is a $D$-dimensional input, and $y$ is the observation. 
Let the dataset $\mathcal{D} = (\mathbf{x}_i, y_i)_{i=1}^n$, where $n$ is the number of observations. 
A GP regression assumes that $\mathbf{f} = f_{1:n}$ is jointly Gaussian, and the observation $y$ is normally distributed given $f$, 
\begin{equation}
\mathbf{f} | \mathbf{x}_1, \dots, \mathbf{x}_n \sim \mathcal{N}(\mathbf{m}, \mathbf{K}),
\end{equation} 
\begin{equation}
\mathbf{y} | \mathbf{f},\sigma^2 \sim \mathcal{N}(\mathbf{f}, \sigma^2 \mathbf{I}),
\end{equation} 
where the element of $\mathbf{m}$ is $\mu_i:=m(\mathbf{x}_i)$ and $\mathbf{K}_{i,j}:= k(\mathbf{x}_i, \mathbf{x}_j)$. 

The covariance kernel $\mathbf{K}$ is a choice of modeling covariance between inputs. At an unknown sampling location $\mathbf{x}$, the predicted response is described by a posterior Gaussian distribution, where the posterior mean is
\begin{equation}
\label{eq:mean}
\mu_n(\mathbf{x}) = \mathbf{m}(\mathbf{x}) + \mathbf{k}(\mathbf{x})^\top (\mathbf{K} + \sigma^2 \mathbf{I})^{-1} (\mathbf{y} - \mathbf{m}),
\end{equation}
and the posterior variance is
\begin{equation}
\label{eq:variance}
\sigma_n^2 = \mathbf{k}(\mathbf{x}, \mathbf{x}) - \mathbf{k}(\mathbf{x})^\top (\mathbf{K}  + \sigma^2 \mathbf{I})^{-1} \mathbf{k}(\mathbf{x}),
\end{equation}
where $k(\mathbf{x})$ is the covariance vector between the query point $\mathbf{x}$ and $\mathbf{x}_{1:n}$. 
The classical GP formulation assumes stationary covariance matrix, which only depends on the distance $r = \lVert \mathbf{x} - \mathbf{x'} \rVert$. 
Several most common kernels for GP include~\cite{shahriari2016taking}
\begin{itemize}
\item $k_{\text{Mat{\'e}rn}1} (\mathbf{x}, \mathbf{x'}) = \theta_0^2 \exp{(-r)}$, 
\item $k_{\text{Mat{\'e}rn}3} (\mathbf{x}, \mathbf{x'}) = \theta_0^2 \exp{(-\sqrt{3}r)} (1+\sqrt{3} r)$, 
\item $k_{\text{Mat{\'e}rn}5} (\mathbf{x}, \mathbf{x'}) = \theta_0^2 \exp{(-\sqrt{5}r)} \left( 1 + \sqrt{5}r + \frac{5}{3}r^2 \right)$, 
\item $k_{\text{sq-exp}} (\mathbf{x}, \mathbf{x'}) = \theta_0^2 \exp{\left(-\frac{1}{2}r^2 \right)}$.
\end{itemize}
The log-likelihood function can be written as
\begin{equation}
\label{eq:posteriorClassicalGP}
\begin{array}{lll}
\log{p(\mathbf{y} | \mathbf{x}_{1:n}, \theta )} &=& - \frac{n}{2} \log{(2\pi)} - \frac{1}{2} \log{| \mathbf{K}^{\theta} + \sigma^2 \mathbf{I} |}  \\
& & - \frac{1}{2} (\mathbf{y} - \mathbf{m})^\top (\mathbf{K}^{\theta} + \sigma^2 \mathbf{I} )^{-1} (\mathbf{y} - \mathbf{m}).
\end{array}
\end{equation}
Optimizing the log marginal likelihood function yields the hyper-parameter $\theta$ at the computational cost of $\mathcal{O}(n^3)$ due to the cost to compute the inverse of the covariance matrix.

We adopt the notation from Qui\~{n}onero-Candela et al. \cite{quinonero2005unifying,quinonero2007approximation} 
but reinstate the non-zero mean assumption\footnote{Qui\~{n}onero-Candela et al. \cite{quinonero2007approximation} assumed $\mathbf{m} = \mathbf{0}$.}. 
Denote the training and testing function values as $\mathbf{f}$ and $\mathbf{f_*}$, respectively, the joint GP prior could be rewritten in a probabilistic manner, i.e.
\begin{equation}
p(\mathbf{f}, \mathbf{f_*}) = \mathcal{N}\left(\begin{bmatrix} \mathbf{m} \\ \mathbf{m} \end{bmatrix}, \begin{bmatrix} \mathbf{K}_{\mathbf{f},\mathbf{f}} & \mathbf{K}_{\mathbf{*},\mathbf{f}} \\ \mathbf{K}_{\mathbf{f},\mathbf{*}} & \mathbf{K}_{\mathbf{*},\mathbf{*}} \end{bmatrix}\right), 
\end{equation}
which leads to the Gaussian predictive distribution by Bayes' rule,
\begin{equation}
\begin{array}{lll}
p(\mathbf{f_*} | \mathbf{y}) &=& \int p(\mathbf{f}, \mathbf{f_*}| \mathbf{y}) d\mathbf{f} \\
&=& \frac{1}{p(\mathbf{y})} \int p(\mathbf{y}|\mathbf{f}) p(\mathbf{f},\mathbf{f_*}) d\mathbf{f} \\
&=& \mathcal{N} (\mathbf{m} + \mathbf{K}_{\mathbf{*},\mathbf{f}}[\mathbf{K}_{\mathbf{f},\mathbf{f}} + \sigma^2 \mathbf{I}]^{-1} (\mathbf{y} - \mathbf{m}),  \\
&& \mathbf{K}_{\mathbf{*},\mathbf{*}} - \mathbf{K}_{\mathbf{*},\mathbf{f}} [\mathbf{K}_{\mathbf{f}, \mathbf{f}} + \sigma^2 \mathbf{I}]^{-1} \mathbf{K}_{\mathbf{f},\mathbf{*}}),
\end{array}
\end{equation}
as any conditional of a Gaussian distribution is also Gaussian \footnote{If $P(\mathbf{f},\mathbf{g}) = \mathcal{N}\left(\begin{bmatrix} a \\ b \end{bmatrix}, \begin{bmatrix}A & C \\ C^\top & B \end{bmatrix}\right)$ then $P(\mathbf{f}| \mathbf{g}) = \mathcal{N}(a + CB^{-1}(y - b), A-CB^{-1}C^\top)$ (cf. Appendix A \cite{quinonero2005unifying}).}. 

Denote $\mu(\mathbf{x})$, $\sigma^2(\mathbf{x})$, and $\theta$ as the posterior mean, the posterior variance, and the hyper-parameters of the objective GP model, respectively. 
$\theta$ is obtained by maximizing the log likelihood estimation over a plausible chosen range. 
Let $\phi(\cdot)$ and $\Phi(\cdot)$ be the standard normal probability distribution function and cumulative distribution function, respectively, and $\mathbf{x}_{\text{best}} = \argmax_{1\leq i \leq n} f(\mathbf{x}_i)$ be the best-so-far sample. 
Rigorously, the acquisition function should be written as $a(\mathbf{x};\{\mathbf{x}_i,y_i \}_{i=1}^n,\theta)$, but for the sake of simplicity, we drop the dependence on the observations and simply write as $a(\mathbf{x})$ and $\mathbb{E}(\cdot)$ is implicitly understood as $\mathbb{E}_{y \sim p(y | \mathcal{D}_n, \mathbf{x})} (\cdot)$.

The PI acquisition function \cite{kushner1964new} is defined as 
  \begin{equation}
  a_{\text{PI}}(\mathbf{x}) = \text{Pr}(y > f(\mathbf{x}_\text{best})) = \E{\mathbbm{1}_{y>f(\mathbf{x}_\text{best})}} = \Phi(\gamma(\mathbf{x})),
  \end{equation}
where
  \begin{equation}
  \label{eq:normalizedZ}
  \gamma(\mathbf{x}) = \frac{\mu(\mathbf{x}) - f(\mathbf{x}_{\text{best}})}{\sigma(\mathbf{x})},
  \end{equation}
indicates the deviation away from the best sample. 
The PI acquisition function is constructed based on the idea of binary utility function, where a unit reward is received if a new best-so-far sample is found, and zero otherwise. 

The EI acquisition function \cite{mockus1975bayesian,mockus1982bayesian,bull2011convergence,snoek2012practical} is defined as 
\begin{equation}
a_{\text{EI}}(\mathbf{x}) = \sigma(\mathbf{x}) \cdot (\gamma(\mathbf{x}) \Phi(\gamma(\mathbf{x})) + \phi(\gamma(\mathbf{x})).
\end{equation}
The EI acquisition is constructed based on an improvement utility function, where the reward is the relative difference if a new best-so-far sample is found, and zero otherwise. 
A closely related generalization of the EI acquisition function, called knowledge-gradient (KG) acquisition function, has been suggested in \cite{scott2011correlated}. Under the assumptions of noise-free and the sampling function is restricted, the EI acquisition function is recovered from the KG acquisition function. 
If the EI acquisition function is rewritten as 
\begin{equation}
\begin{array}{lll}
a_\text{EI}(\mathbf{x}) &=& \E{\max\left(y, f(\mathbf{x}_\text{best})\right) - f(\mathbf{x}_\text{best})} \\
&=& \E{\max(y - f(\mathbf{x}_\text{best}, 0)} \\
&=& \E{(y - f(\mathbf{x}_\text{best})^+},
\end{array}
\end{equation} 
then the KG acquisition function is expressed as 
\begin{equation}
a_\text{KG}(\mathbf{x}) = \E{ \max \mu_{n+1}(\mathbf{x}) | \mathbf{x}_{n+1} = \mathbf{x}} - \max(\mu_n(\mathbf{x}))
\end{equation}
for one-step look-ahead acquisition function. 

The UCB acquisition function \cite{auer2002using,srinivas2009gaussian,srinivas2012information} is defined as
\begin{equation}
a_{\text{UCB}}(\mathbf{x}) = \mu(\mathbf{x}) + \kappa \sigma(\mathbf{x}),
\end{equation}
where $\kappa$ is a hyper-parameter describing the acquisition exploitation-exploration balance. 
Here, we adopt the $\kappa$ computation from Daniel et al. \cite{daniel2014active}, where
\begin{equation}
\kappa = \sqrt{\nu \gamma_n},\quad \nu = 1, \quad \gamma_n = 2\log{\left(\frac{n^{D/2 + 2}\pi^2}{3\delta} \right)},
\end{equation}
and $D$ is the dimensionality of the problem, and $\delta \in (0,1)$ \cite{srinivas2012information}.

\subsection{Sparse Gaussian process}
We briefly review different approaches presented by Qui\~{n}onero-Candela et al. \cite{quinonero2005unifying,quinonero2007approximation} before adopting the fully independent condition (FIC) as the underlying sparse GP for BO, which approximates the covariance matrix using the low-rank Nystr\"{o}m approximation $\mathbf{K} \approx \widetilde{\mathbf{K}} = \mathbf{K}_{n\times m} \mathbf{K}_{m \times m}^{-1} \mathbf{K}_{m \times n}$ (cf. Section 8.1 \cite{rasmussen2006gaussian}) and scales as $\mathcal{O}(nm^2 + m^3)$ instead of $\mathcal{O}(n^3)$. For $n \gg m$, this method scales as $\mathcal{O}(nm^2)$. 

Qui\~{n}onero-Candela et al. \cite{quinonero2007approximation} provided an extensive discussion and theoretical comparison between different sparse GP approaches. 
Following Qui\~{n}onero-Candela et al. \cite{quinonero2005unifying,quinonero2007approximation} and Chalupka et al.\cite{chalupka2013framework}, Vanhatalo et al. \cite{vanhatalo2012bayesian,vanhatalo2013gpstuff}, we introduce the inducing variables $\mathbf{u}$ as values of the GP (as also $\mathbf{f}$ and $\mathbf{f_*}$) and marginalize $\mathbf{u}$ in the predictive distribution as
\begin{equation}
p(\mathbf{f_*}, \mathbf{f}) = \int p(\mathbf{f_*}, f, \mathbf{u}) d\mathbf{u} = \int p(\mathbf{f_*}, \mathbf{f} | \mathbf{u}) p(\mathbf{u}) d\mathbf{u},
\end{equation}
where $p(\mathbf{u}) = \mathcal{N}(\mathbf{m}, \mathbf{K}_{\mathbf{u},\mathbf{u}})$. In sparse GP approximations, the joint prior is approximated by assuming that $\mathbf{f_*}$ and $\mathbf{f}$ are conditionally independent given $\mathbf{u}$, i.e.
\begin{equation}
\label{eq:ConditionallyIndependent}
p(\mathbf{f_*}, \mathbf{f}) \approx q(\mathbf{f_*}, \mathbf{f}) = \int q(\mathbf{f_*} | \mathbf{u}) q(\mathbf{f}| \mathbf{u}) p(\mathbf{u}) d\mathbf{u}.
\end{equation}
In the noise-free settings, the training conditional is
\begin{equation}
p(\mathbf{f}| \mathbf{u}) = \mathcal{N}(\mathbf{m} + \mathbf{K}_{\mathbf{f},\mathbf{u}} \mathbf{K}^{-1}_{\mathbf{u}, \mathbf{u}} (\mathbf{u} - \mathbf{m}) ,\ \mathbf{K}_{\mathbf{f}, \mathbf{f}} - \mathbf{Q}_{\mathbf{f}, \mathbf{f}})
\end{equation}
whereas the test conditional is
\begin{equation}
p(\mathbf{f_*}| \mathbf{u}) = \mathcal{N}(\mathbf{m} + \mathbf{K}_{\mathbf{*},\mathbf{u}} \mathbf{K}^{-1}_{\mathbf{u}, \mathbf{u}} (\mathbf{u} - \mathbf{m}) ,\ \mathbf{K}_{\mathbf{*}, \mathbf{*}} - \mathbf{Q}_{\mathbf{*}, \mathbf{*}}),
\end{equation}
where $\mathbf{Q}_{\cdot, \cdot}$ is defined as 
\begin{equation}
\label{eq:Qdef}
\mathbf{Q}_{\mathbf{a}, \mathbf{b}} := \mathbf{K}_{\mathbf{a},\mathbf{u}} \mathbf{K}^{-1}_{\mathbf{u}, \mathbf{u}} \mathbf{K}_{\mathbf{u},\mathbf{b}}. 
\end{equation}
It is noted that $\mathbf{K}_{\mathbf{*},\mathbf{*}} - \mathbf{Q}_{\mathbf{*},\mathbf{*}} = \mathbf{K}_{\mathbf{*},\mathbf{*}} - \mathbf{K}_{\mathbf{*},\mathbf{u}} \mathbf{K}^{-1}_{\mathbf{u}, \mathbf{u}} \mathbf{K}_{\mathbf{u},\mathbf{*}}$ is a Schur complement of $\mathbf{K}$, i.e. $\mathbf{K} / \mathbf{K}_{\mathbf{u},\mathbf{u}}$ (cf. Rasmussen \cite{rasmussen2006gaussian} Section 8.6). 
The exact likelihood and inducing prior remain the same, i.e.
\begin{equation}
p(\mathbf{y}| \mathbf{f}) = \mathcal{N}(\mathbf{f}, \sigma^2 \mathbf{I}),
\end{equation}
and
\begin{equation}
p(\mathbf{u}) = \mathcal{N}(\mathbf{m}, \mathbf{K}_{\mathbf{u},\mathbf{u}}).
\end{equation}

\subsubsection{Subset of regressors (Deterministic Inducing Conditional approximation)}


The Subset of Regressors (SoR) approximate conditional distributions are given by
\begin{equation}
q_{\text{SoR}}(\mathbf{f}|\mathbf{u}) = \mathcal{N}(\mathbf{m} + \mathbf{K}_{\mathbf{f},\mathbf{u}}\mathbf{K}^{-1}_{\mathbf{u},\mathbf{u}} (\mathbf{u} - \mathbf{m}), 0),
\end{equation}
and
\begin{equation}
q_{\text{SoR}}(\mathbf{f_*}|\mathbf{u}) = \mathcal{N}(\mathbf{m} + \mathbf{K}_{\mathbf{f},\mathbf{u}}\mathbf{K}^{-1}_{\mathbf{*},\mathbf{u}} (\mathbf{u} - \mathbf{m}), 0),
\end{equation}
with zero conditional covariance. 
Following Equation \ref{eq:ConditionallyIndependent}, the effective prior is therefore
\begin{equation}
q_{\text{SoR}}(\mathbf{f}, \mathbf{f_*}) = \mathcal{N}\left(\begin{bmatrix}\mathbf{m} \\ \mathbf{m} \end{bmatrix}, \begin{bmatrix} \mathbf{Q}_{\mathbf{f},\mathbf{f}} & \mathbf{Q}_{\mathbf{f}, \mathbf{*}} \\ \mathbf{Q}_{\mathbf{*},\mathbf{f}} & \mathbf{Q}_{\mathbf{*}, \mathbf{*}} \end{bmatrix}\right),
\end{equation}
where $\mathbf{Q}_{\cdot, \cdot}$ is defined in Equation \ref{eq:Qdef}. 
The predictive distribution is computed as
\begin{equation}
\begin{array}{lll}
q_{\text{SoR}}(\mathbf{f_*} | \mathbf{y}) &=& \mathcal{N}(\mathbf{m} + \mathbf{Q}_{\mathbf{*},\mathbf{f}}[\mathbf{Q}_{\mathbf{f},\mathbf{f}} + \sigma^2 \mathbf{I}]^{-1} (\mathbf{y} - \mathbf{m}), \\ 
&& \mathbf{Q}_{\mathbf{*},\mathbf{*}} - [\mathbf{Q}_{\mathbf{*},\mathbf{f}} + \sigma^2 \mathbf{I}]^{-1} \mathbf{Q}_{\mathbf{f},\mathbf{*}} ) \\
&=& \mathcal{N}(\mathbf{m} + \sigma^{-2} \mathbf{K}_{\mathbf{*},\mathbf{u}} \Sigma \mathbf{K}_{\mathbf{u},\mathbf{f}} (\mathbf{y} - \mathbf{m}),\ \mathbf{K}_{\mathbf{*},\mathbf{u}} \Sigma \mathbf{K}_{\mathbf{u},\mathbf{*}}),
\end{array}
\end{equation}
where
\begin{equation}
\label{eq:Sigma}
\Sigma = [\sigma^2 \mathbf{K}_{\mathbf{u},\mathbf{f}} \mathbf{K}_{\mathbf{f},\mathbf{u}} + \mathbf{K}_{\mathbf{u},\mathbf{u}}]^{-1}. 
\end{equation}

\subsubsection{Deterministic Training Conditional approximation (DTC)}

Let the projection $\mathbf{f} = \mathbf{m} + \mathbf{K}_{\mathbf{f},\mathbf{u}} \mathbf{K}_{\mathbf{u},\mathbf{u}}^{-1} (\mathbf{u} - \mathbf{m})$, then
\begin{equation}
p(\mathbf{y}|\mathbf{f}) \approx q(\mathbf{y}|\mathbf{u}) = \mathcal{N}(\mathbf{m} + \mathbf{K}_{\mathbf{f},\mathbf{u}} \mathbf{K}_{\mathbf{u},\mathbf{u}}^{-1} (\mathbf{u} - \mathbf{m}), \sigma^2 \mathbf{I})
\end{equation}
By imposing a deterministic training conditional and retaining the usual likelihood and the exact test conditional,
\begin{equation}
q_{\text{DTC}} (\mathbf{f} | \mathbf{u}) = \mathcal{N}(\mathbf{m} + \mathbf{K}_{\mathbf{f},\mathbf{u}} \mathbf{K}_{\mathbf{u},\mathbf{u}}^{-1} (\mathbf{u} - \mathbf{m}), 0)
\end{equation}
and
\begin{equation}
q_{\text{DTC}}(\mathbf{f_*} | \mathbf{u}) = p(\mathbf{f_*} | \mathbf{u}),
\end{equation}
under the joint prior implied by DTC as
\begin{equation}
q_\text{DTC}(\mathbf{f},\mathbf{f_*}) = \mathcal{N}\left(\begin{bmatrix} \mathbf{m} \\ \mathbf{m} \end{bmatrix}, \begin{bmatrix} \mathbf{Q}_{\mathbf{f},\mathbf{f}} & \mathbf{Q}_{\mathbf{f},\mathbf{*}} \\ \mathbf{Q}_{\mathbf{*},\mathbf{f}} & \mathbf{K}_{\mathbf{*},\mathbf{*}} \end{bmatrix}\right),
\end{equation}
the predictive distribution is given by
\begin{equation}
\begin{array}{lll}
q_{\text{DTC}}(\mathbf{f_*} | \mathbf{y}) &=& \mathcal{N}(\mathbf{m} + \mathbf{Q}_{\mathbf{*}, \mathbf{f}} [\mathbf{Q}_{\mathbf{f},\mathbf{f}} + \sigma^2 \mathbf{I}]^{-1} (\mathbf{y} - \mathbf{m}), \\
&& \mathbf{K}_{\mathbf{*},\mathbf{*}} - \mathbf{Q}_{\mathbf{*},\mathbf{f}} [\mathbf{Q}_{\mathbf{f},\mathbf{f}} + \sigma^2 \mathbf{I}]^{-1} \mathbf{Q}_{\mathbf{f}, \mathbf{*}}) \\
\end{array},
\end{equation}
where $\Sigma$ is defined in Equation \ref{eq:Sigma}. It is noted that $\mathbf{K}_{\mathbf{*},\mathbf{*}} - \mathbf{Q}_{\mathbf{*},\mathbf{*}}$ is positive semidefinite and that the variance of DTC is always larger or equal to that of SoR and the DTC does not correspond exactly to a GP.

\subsubsection{Partially Independent (Training) Conditional (PIC)}
The Partially Independent (Training) Conditional (PIC) approximations assumes the training conditional 
\begin{equation}
q_{\text{PIC}}(\mathbf{f} | \mathbf{u}) = \mathcal{N}(\mathbf{m} + \mathbf{K}_{\mathbf{f},\mathbf{u}} \mathbf{K}_{\mathbf{u},\mathbf{u}}^{-1} (\mathbf{u} - \mathbf{m}), \text{BlockDiag}[\mathbf{K}_{\mathbf{f},\mathbf{f}} - \mathbf{Q}_{\mathbf{f},\mathbf{f}}]),
\end{equation}
and testing conditional 
\begin{equation}
q_\text{PIC}(\mathbf{f_*} | \mathbf{u} ) = p(\mathbf{f_*} | \mathbf{u}),
\end{equation}
where $\text{BlockDiag}[\cdot]$ is a block-diagonal matrix. The effective prior implied by PIC is
\begin{equation}
q_\text{PIC}(\mathbf{f}, \mathbf{f_*}) = \mathcal{N}\left( \begin{bmatrix} \mathbf{m} \\ \mathbf{m} \end{bmatrix}, \begin{bmatrix} \mathbf{Q}_{\mathbf{f},\mathbf{f}} - \text{BlockDiag}[\mathbf{Q}_{\mathbf{f},\mathbf{f}} - \mathbf{K}_{\mathbf{f},\mathbf{f}}] & \mathbf{Q}_{\mathbf{f},\mathbf{*}} \\ \mathbf{Q}_{\mathbf{*},\mathbf{f}} & \mathbf{K}_{\mathbf{*},\mathbf{*}} \end{bmatrix} \right),
\end{equation}
which leads to the predictive distribution of
\begin{equation}
\label{eq:posteriorPIC}
\begin{array}{lll}
q_{\text{PIC}}(\mathbf{f_*} | \mathbf{y}) &=& \mathcal{N}(\mathbf{m} + \mathbf{Q}_{\mathbf{*}, \mathbf{f}} [\mathbf{Q}_{\mathbf{f},\mathbf{f}} + \Lambda]^{-1} (\mathbf{y} - \mathbf{m}), \\
&& \mathbf{K}_{\mathbf{*},\mathbf{*}} - \mathbf{Q}_{\mathbf{*},\mathbf{f}} [\mathbf{Q}_{\mathbf{f},\mathbf{f}} + \Lambda]^{-1} \mathbf{Q}_{\mathbf{f}, \mathbf{*}}) \\ 
&=& \mathcal{N}(\mathbf{m} + \mathbf{K}_{\mathbf{*}, \mathbf{u}} \Sigma \mathbf{K}_{\mathbf{u},\mathbf{f}} \Lambda^{-1} (\mathbf{y} - \mathbf{m}), \\
&& \mathbf{K}_{\mathbf{*},\mathbf{*}} - \mathbf{Q}_{\mathbf{*},\mathbf{*}} + \mathbf{K}_{\mathbf{*},\mathbf{u}} \Sigma \mathbf{K}_{\mathbf{u}, \mathbf{*}}) \\
\end{array},
\end{equation}
where $\Sigma = [\mathbf{K}_{\mathbf{u},\mathbf{u}} + \mathbf{K}_{\mathbf{u},\mathbf{f}} \Lambda^{-1} \mathbf{K}_{\mathbf{f},\mathbf{u}}]^{-1}$ and $\Lambda = \text{BlockDiag}[\mathbf{K}_{\mathbf{f},\mathbf{f}} - \mathbf{Q}_{\mathbf{f},\mathbf{f}} + \sigma^2 \mathbf{I}]$ by utilizing the Sherman-Morrison-Woodbury formula. 

\subsubsection{Fully Independent (Training) Conditional approximation (FIC)}
\label{subsubsec:FIC-SparseGP}

If the $\text{BlockDiag}[\cdot]$ is replaced by $\text{Diag}[\cdot]$, then the PIC prior becomes the Fully Independent (Training) Conditional (FIC) prior
\begin{equation}
q_\text{FIC}(\mathbf{f}, \mathbf{f_*}) = \mathcal{N}\left( \begin{bmatrix} \mathbf{m} \\ \mathbf{m} \end{bmatrix}, \begin{bmatrix} \mathbf{Q}_{\mathbf{f},\mathbf{f}} - \text{Diag}[\mathbf{Q}_{\mathbf{f},\mathbf{f}} - \mathbf{K}_{\mathbf{f},\mathbf{f}}] & \mathbf{Q}_{\mathbf{f},\mathbf{*}} \\ \mathbf{Q}_{\mathbf{*},\mathbf{f}} & \mathbf{K}_{\mathbf{*},\mathbf{*}} \end{bmatrix} \right).
\end{equation}
which leads to the predictive distribution of
\begin{equation}
\begin{array}{lll}
q_{\text{FIC}}(\mathbf{f_*} | \mathbf{y}) &=& \mathcal{N}(\mathbf{m} + \mathbf{K}_{\mathbf{*}, \mathbf{u}} \Sigma \mathbf{K}_{\mathbf{u},\mathbf{f}} \Lambda^{-1} (\mathbf{y} - \mathbf{m}), \\
&& \mathbf{K}_{\mathbf{*},\mathbf{*}} - \mathbf{Q}_{\mathbf{*},\mathbf{*}} + \mathbf{K}_{\mathbf{*},\mathbf{u}} \Sigma \mathbf{K}_{\mathbf{u}, \mathbf{*}}) \\
\end{array},
\end{equation}
where $\Sigma = [\mathbf{K}_{\mathbf{u},\mathbf{u}} + \mathbf{K}_{\mathbf{u},\mathbf{f}} \Lambda^{-1} \mathbf{K}_{\mathbf{f},\mathbf{u}}]^{-1}$ and $\Lambda = \text{Diag}[\mathbf{K}_{\mathbf{f},\mathbf{f}} - \mathbf{Q}_{\mathbf{f},\mathbf{f}} + \sigma^2 \mathbf{I}]$, which is nearly identical with Equation \ref{eq:posteriorPIC} except for the definition of $\Lambda$. With this formulation, we recover the SSGP approach proposed by Snelson and Ghahramani \cite{snelson2005sparse}.

\subsubsection{Training sparse GP}

Following Qui\~{n}onero-Candela et al. \cite{quinonero2005unifying,quinonero2007approximation}, the marginal likelihood conditioned on the inducing inputs is given by
\begin{equation}
\begin{array}{lll}
q(\mathbf{y}| \mathbf{X_u}) &=& \int \int p(\mathbf{y} | \mathbf{f}) q(\mathbf{f} | \mathbf{u}) p(\mathbf{u} | \mathbf{X_u}) d\mathbf{u} d\mathbf{f} \\
&=& \int p(\mathbf{y} |\mathbf{f}) q(\mathbf{f}| \mathbf{X_u})d\mathbf{f}.
\end{array}
\end{equation}
Using the corresponding definitions of $\Lambda$ (depending on the choice of sparse GP models\footnote{$\Lambda_{\text{SoR}} = \Lambda_{\text{DTC}} = \sigma^2 \mathbf{I}$, $\Lambda_\text{PITC}= \text{BlockDiag}[\mathbf{K}_{\mathbf{f},\mathbf{f}} - \mathbf{Q}_{\mathbf{f},\mathbf{f}}] + \sigma^2 \mathbf{I}$, $\Lambda_\text{FIC}= \text{Diag}[\mathbf{K}_{\mathbf{f},\mathbf{f}} - \mathbf{Q}_{\mathbf{f},\mathbf{f}}] + \sigma^2 \mathbf{I}$.}), the log marginal likelihood is
\begin{equation}
\label{eq:LogLikelihoodSparseGP}
\begin{array}{lll}
\log q(\mathbf{y}|\mathbf{X_u}, \theta) &=& - \frac{n}{2} \log(2 \pi) - \frac{1}{2} \log|\mathbf{Q}_{\mathbf{f},\mathbf{f}}  + \Lambda| \\
&&- \frac{1}{2} (\mathbf{y} - \mathbf{m})^\top [\mathbf{Q}_{\mathbf{f},\mathbf{f}} + \Lambda]^{-1} (\mathbf{y} - \mathbf{m}),
\end{array}
\end{equation}
which is strikingly similar to the case of classical GP as in Equation \ref{eq:posteriorClassicalGP}.
Invoking Sherman-Morrison-Woodbury formula\footnote{\label{fn:SMWidentity}$(A+U B V^\top)^{-1} = A^{-1} - A^{-1} U (B^{-1} + V^\top A^{-1} U)^{-1} V^\top A^{-1}$.} and realizing that $\mathbf{Q}_{\mathbf{f},\mathbf{f}} = \mathbf{K}_{\mathbf{f},\mathbf{u}} \mathbf{K}_{\mathbf{u},\mathbf{u}}^{-1} \mathbf{K}_{\mathbf{u},\mathbf{u}}$ by Equation \ref{eq:Qdef}, 
then
\begin{equation}
\begin{array}{lll}
[\mathbf{Q}_{\mathbf{f},\mathbf{f}} + \Lambda]^{-1} &=& [\Lambda + \mathbf{K}_{\mathbf{f},\mathbf{u}} \mathbf{K}_{\mathbf{u},\mathbf{u}}^{-1} \mathbf{K}_{\mathbf{u},\mathbf{f}}]^{-1} \\
&=& \Lambda^{-1}  - \Lambda^{-1} \mathbf{K}_{\mathbf{f},\mathbf{u}}[\mathbf{K}_{\mathbf{u},\mathbf{u}} + \mathbf{K}_{\mathbf{u},\mathbf{f}} \Lambda^{-1} \mathbf{K}_{\mathbf{f},\mathbf{u}}]^{-1} \\
&& \quad \quad \mathbf{K}_{\mathbf{u},\mathbf{f}} \Lambda^{-1}.
\end{array}
\end{equation}
Optimizing the log marginal likelihood yields the hyper-parameters $\theta$ as in the case of the classical GP. 

An alternative training approach is to maximize the (variational) evidence lower bound (ELBO) \cite{titsias2009variational,titsias2009variational2} of the true log marginal likelihood in Equation \ref{eq:LogLikelihoodSparseGP} as
\begin{equation}
\label{eq:ELBOSparseGP}
\begin{array}{lll}
ELBO(\theta, \mathbf{X_u}) &=& \log[ \mathcal{N}(\mathbf{y} | \mathbf{m}, \sigma^2 \mathbf{I} + \mathbf{Q}_{\mathbf{f},\mathbf{f}})] \\
&& - \frac{1}{2\sigma^2} \text{Tr}\left[\mathbf{K}_{\mathbf{f},\mathbf{f}} - \mathbf{K}_{\mathbf{f},\mathbf{u}} \mathbf{K}_{\mathbf{u},\mathbf{u}}^{-1} \mathbf{K}_{\mathbf{u},\mathbf{f}}\right]
\end{array}
\end{equation}
In either approach, the overall computational cost is $\mathcal{O}(nm^2)$ to compute $\Sigma$ \cite{candela2004learning}, $\mathcal{O}(m)$ and $\mathcal{O}(m^2)$ to compute the posterior mean and variance, respectively \cite{quinonero2007approximation}. This is to contrast with the complexity of $\mathcal{O}\left(\frac{1}{3}n^3\right)$ for Cholesky, $\mathcal{O}\left(\frac{2}{3}n^3\right)$ for LU, and $\mathcal{O}\left(\frac{4}{3}n^3\right)$ for QR decomposition algorithms (cf. Table C.2 \cite{higham2008functions}).

\subsection{High-dimensional Bayesian optimization via random embeddings}

We adopt the random embeddings proposed by Wang et al. \cite{wang2013bayesian,wang2016bayesian} for reducing the dimensionality by rotating the input with a random matrix $\mathbf{A}$. Of similar yet distinct idea is the active subspace method by Constantine et al \cite{constantine2014active,constantine2015active}. It is worthy to mention that there has been some significant efforts to reduce the dimensionality in Bayesian optimization using active subspace \cite{tripathy2016gaussian,gautier2020fully}. However, the true difficulty lies in the approximation of the high-dimensional gradients in the classical GP approach, which is originally gradient-free, in order to obtain an optimal rotation matrix on the Stiefel \cite{tripathy2016gaussian,gautier2020fully} or Grassmann manifolds\footnote{If $\mathbf{A} \in \mathbb{R}^{D\times d}$ with $a_{ij} \overset{\text{i.i.d}}{\sim} \mathcal{N}(0,1)$, then $\mathbf{Q} = \mathbf{A}(\mathbf{A}^\top \mathbf{A})^{-1/2}$ (i.e. $\mathbf{A} = \mathbf{Q} \mathbf{R}$) is uniformly distributed on the Stiefel manifold $V_{d,D}$ (cf Theorem 2.2.1 \cite{chikuse2012statistics}).} \cite{seshadri2019dimension,hokanson2018data} in the active subspace approach. This, unnecessarily, complicates the optimization problem (by assigning the additional task of discovering the active subspace, which is usually treated as a constrained manifold optimization problem) that is already challenging in high-dimensional space.

Random embeddings \cite{wang2013bayesian,wang2016bayesian}, on another hand, do not suffer from the task of obtaining the active subspace. It turns out that one does not completely require \textit{a priori} knowledge of the active subspace for optimization purposes. The random embeddings approach completely ignores the task of discovering the active subspace, but instead focuses on exploiting the existence of an (unknown) active subspace to reduce the dimensionality of the problem. As explained by Letham et al. \cite{letham2020re}, random embeddings comes with a strong theoretical guarantee (Theorem \ref{thm:ProbabilityGuarantee}) because by the Johnson-Lindenstrauss lemma \cite{johnson1984extensions}, random projections can approximately preserve $\ell^2$-distance preserved, up to a $(1\pm \varepsilon)$-factor \cite{nayebi2019framework}, without requiring data to learn the embedding. It is noted that alternative embedding approach is also discussed in HeSBO by Nayebi et al \cite{nayebi2019framework} based on hashing.

\begin{algorithm}[!htbp]
\caption{REMBO algorithm \cite{wang2016bayesian} with deviation from BO highlighted.}
\label{alg:rembo}

\begin{algorithmic}[1]

\State \blue{draw a random matrix $\mathbf{A} \in \mathbb{R}^{D\times d}: a_{ij} \overset{\text{i.i.d}}{\sim} \mathcal{N}(0,1)$}
\State \blue{choose the bounded region set $\mathcal{Z} \subset \mathbb{R}^d$}
\State $\mathcal{D}_0 \gets \varnothing$
\For{$i=1,2,\cdots$}
\State locate next sampling point $\color{blue}{\mathbf{z}_{i+1} \gets \argmax_{\mathbf{z} \in \mathcal{Z}} a(\mathbf{z}) \in \mathbb{R}^d}$
\State query $\mathcal{D}_{i+1} \gets \mathcal{D}_i \cup \color{blue}{ \{ z_{i+1}, f(p_{\mathcal{X}} (\mathbf{A} z_{i+1})) \} }$
\State update GP
\EndFor

\end{algorithmic}
\end{algorithm}

Following Wang et al. \cite{wang2013bayesian,wang2016bayesian}, we briefly summarize REMBO formulation in Algorithm \ref{alg:rembo}.

\begin{definition}
A function $f:\mathbb{R}^D \to \mathbb{R}$ is said to have \textbf{effective dimensionality} $d_e$ with $d_e \leq D$ if
\begin{itemize}
\item there exists a linear subspace $\mathcal{T}$ of dimension $d_e$ such that for all $\mathbf{x}_{\top} \in \mathcal{T} \subset \mathbb{R}^D$ and $\mathbf{x}_{\perp} \in \mathcal{T}^{\perp} \subset \mathbb{R}^D$, where $\mathcal{T}^{\perp}$ denotes the orthogonal complement of $\mathcal{T}$; and 
\item $d_e$ is the smallest integer with this property.
\end{itemize}
$\mathcal{T}$ is called the \textbf{effective subspace} of $f$ and $\mathcal{T}^{\perp}$ the \textbf{constant subspace}.
\end{definition}

\begin{theorem}[cf. Theorem 2 \cite{wang2016bayesian}]
\label{thm:REMBO}
Assume that we are given a function $f:\mathbb{R}^D \to \mathbb{R}$ with effective dimensionality $d_e$ and a random matrix $\mathbf{A} \in \mathbb{R}^{D\times d}$ where $a_{ij} \overset{\text{i.i.d}}{\sim} \mathcal{N}(0,1)$ and $d \geq d_e$. Then, with probability 1, for any $\mathbf{x} \in \mathbb{R}^D$, there exists a $\mathbf{z} \in \mathbb{R}^d$ such that $f(\mathbf{x}) = f(\mathbf{A} \mathbf{z})$. 
\end{theorem}

A significant corollary of Theorem \ref{thm:REMBO} is the existence of $\mathbf{z^*} \in \mathbb{R}^d$ such that $f(\mathbf{x^*}) = f(\mathbf{A} \mathbf{z^*})$, where $\mathbf{x^*} = \argmax_{\mathbf{x} \in \mathcal{X}} f(\mathbf{x})$ is the global optimum. Therefore, instead of optimizing $f(\mathbf{x})$ in the high-dimensional space $\mathcal{X} \subset \mathbb{R}^{D}$, one could optimize the function $f(\mathbf{A} \mathbf{z})$ in the low-dimensional space $\mathcal{Z} \subset \mathbb{R}^d$, which is the main idea of REMBO \cite{wang2016bayesian}. To resolve the topology and magnitude difference between $\mathcal{X}$ and $\mathbf{A} \mathcal{Z}$, one actually optimizes the function $f(p_{\mathcal{X}} (\mathbf{A} \mathbf{z}))$, where $p_{\mathcal{X}}(\mathbf{A} \mathbf{z}) = \argmin_{\mathbf{x} \in \mathcal{X}} \lVert \mathbf{x} - \mathbf{A} \mathbf{z}  \rVert_2$ is projection operator. This also suggests choosing $\mathcal{Z} = \left[ - \frac{1}{\varepsilon} \max\{\log(d_e), 1\}, \frac{1}{\varepsilon} \max\{\log(d_e), 1\} \right]^{d_e}$, as described in Theorem \ref{thm:ProbabilityGuarantee}. We follow Wang et al. \cite{wang2016bayesian} to choose $\varepsilon = \frac{\log d}{d}$, which implies $\mathcal{Z} \subseteq [-\sqrt{d}, \sqrt{d}]^d$, conservatively. 

\begin{theorem}[cf. Theorem 3 \cite{wang2016bayesian}]
\label{thm:ProbabilityGuarantee}
Suppose we want to optimize a function $f: \mathbb{R}^{D} \rightarrow \mathbb{R}$ with effective dimension $d_e \leq d$ subject to the box constraint $\mathcal{X} \subset \mathbb{R}^D$, where $\mathcal{X}$ is centered around $\mathbf{0}$. 
Suppose further that the effective subspace $\mathcal{T}$ of $f$ is such that $\mathcal{T}$ is the span of $d_e$ basis vectors, and let $\mathbf{x}^{*}_\top \in \cal{T} \cap \mathcal{X}$ be an optimizer of $f$ inside $\mathcal{T}$. 
If $\mathbf{A}$ is a $D\times d$ random matrix with independent standard Gaussian entries,
there exists an optimizer $\mathbf{z}^* \in \mathbb{R}^{d}$ such that $f(\mathbf{A}\mathbf{z}^*) = f(\mathbf{x}^*_\top)$ and $\|\mathbf{z}^*\|_2 \leq \frac{\sqrt{d_e}}{\epsilon}\|\mathbf{x}^{*}_\top\|_2$ with probability at least $1-\epsilon$.
\end{theorem}

Technical proof of Theorem \ref{thm:ProbabilityGuarantee} relies on Edelman's theorem (cf. Theorem 3.4~\cite{sankar2006smoothed} and the original paper~\cite{edelman1988eigenvalues}), which is built upon the framework of Gaussian random projection. This work would be incomplete without stating the Johnson-Lindenstrauss lemma in this paper's notation. 

\begin{theorem}[Johnson-Lindenstrauss lemma (cf. Lemma 15~\cite{mahoney2016lecture})]
\label{}
Given $n$ points $\{\mathbf{x}_i\}_{i=1}^n$, each of which is in $\mathbb{R}^D$, $\mathbf{A} \in \mathbb{R}^{D \times d}$ be such that $a_{ij} \overset{\text{i.i.d}}{\sim} \frac{1}{\sqrt{d}} \mathcal{N}(0,1)$, and let $\mathbf{z} \in \mathbb{R}^d$ defined as $\mathbf{z} =  \mathbf{A}^\top \mathbf{x}$. Then, if $d \geq \frac{9 \log n}{\varepsilon^2 - \varepsilon^3}$, for some $\varepsilon \in \left(0, \frac{1}{2} \right)$, then with probability at least $\frac{1}{2}$, all pairwise distances are preserved, i.e. for all $i, j$, we have
\begin{equation}
(1 - \varepsilon) \lVert \mathbf{x}_i - \mathbf{x}_j \ \rVert_2^2 \leq \lVert \mathbf{z}_i - \mathbf{z}_j \rVert_2^2 \leq (1 + \varepsilon) \lVert \mathbf{x}_i - \mathbf{x}_j \ \rVert_2^2
\end{equation}
\end{theorem}


\subsection{Asynchronous parallelism}

\begin{figure*}[!htbp]
\centering
\subcaptionbox{Batch-sequential parallel.
\label{fig:batchSequentialParallel}
}
  [.455\linewidth]{\includegraphics[width=0.455\textwidth, keepaspectratio]{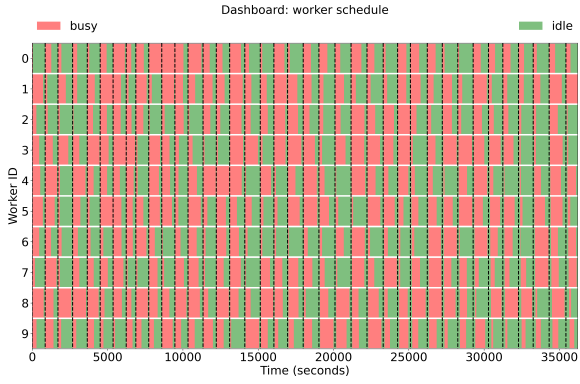}}
\hfill
\subcaptionbox{Asynchronous parallel.
\label{fig:asynchronousParallel}
}
  [.455\linewidth]{\includegraphics[width=0.455\textwidth, keepaspectratio]{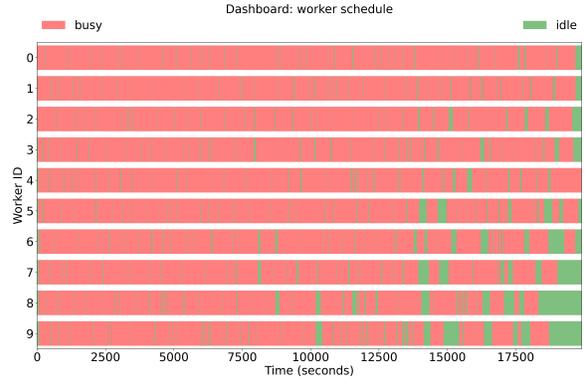}}
\caption{Difference between batch-sequential and asynchronous parallelizations. 
In batch-sequential parallelization \cite{tran2019pbo} (Figure \ref{fig:batchSequentialParallel}), all workers need to wait until the batch is finished before moving on to the next batch. 
In asynchronous parallelization \cite{tran2020aphbo} (Figure \ref{fig:asynchronousParallel}), all the workers receive the most up-to-date information and work independently with each other. 
Asynchronous parallelization is more efficient than batch-sequential parallelization because it reduces idle time for workers. 
}
\label{fig:cropped.asyncParBOScheme}
\end{figure*}

\begin{algorithm*}[!htbp]
\caption{GP-UCB-PE \cite{contal2013parallel}.}
\label{alg:gp-ucb-pe}

\begin{algorithmic}[1]

\For{$i=0, 1, \cdots$}
\State update GP
\State locate next sampling point $\mathbf{x}_{i+1} \gets \argmax_{\mathbf{x} \in \mathcal{X}} a(\mathbf{x})$
\For{$j=0,\cdots, \text{BatchSize} - 1$}
\State update GP
\State locate next sampling point $\mathbf{x}_{i+1} \gets \argmax_{\mathbf{x} \in \mathcal{X}} \sigma^2(\mathbf{x})$
\EndFor
\EndFor

\end{algorithmic}

\end{algorithm*}

We adopt the asynchronous parallel BO framework in Tran et al. \cite{tran2020aphbo}, which leverages the idea of batch-sequential parallel BO in Tran et al. pBO-2GP-3B \cite{tran2019pbo} and fully desynchronizes the workers across the batch. However, in this work, to demonstrate the idea of Scalable$^3$-BO, we exclude the GP-Hedge algorithm \cite{hoffman2011portfolio}, as well as the constraint module, and focus on the numerical Scalable$^3$-BO method with different acquisition functions. Figure \ref{fig:cropped.asyncParBOScheme} highlights the difference between batch-sequential parallel and asynchronous parallel BO methods, where in the asynchronous parallel, workers do not need to wait for others to finish their jobs in order to launch their next jobs. In the nutshell, the asynchronous feature is implemented by constantly checking if any job has been finished, and subsequently assigns idle worker(s) with corresponding task and batch, whether it is to exploit/explore or purely explore. The asynchronous parallel feature is implemented on SLURM and PBS schedulers, as well as regular multi-core workstations. 

The main idea of parallel BO methods proposed by Tran et al. \cite{tran2019pbo,tran2020aphbo} is built on top of previous works of GP-BUCB-AUCB by Desautels et al. \cite{desautels2014parallelizing} and GP-UCB-PE by Contal et al \cite{contal2013parallel}., where the underlying GP is temporarily ``hallucinated'' by the GP posterior mean. The hallucination is removed and the underlying GP is updated whenever the true response is available. To include constraints, another GP or probabilistic binary classifier is used to distinguish feasible versus infeasible inputs. However, in the scope of this paper, constraints are not considered, and thus, there are only two batches: one batch for maximizing the acquisition function, whereas another batch for maximizing the GP posterior variance. Algorithm \ref{alg:gp-ucb-pe} summarizes GP-UCB-PE algorithm proposed by Contal et al. \cite{contal2013parallel}, where the second \serif{For} loop iterates through the batch and wait until the last worker finishes its job, as illustrated in Figure \ref{fig:batchSequentialParallel}.

\subsection{Proposed method: Scalable$^3$-Bayesian-Optimization}

\begin{algorithm*}[!htbp]
\caption{Scalable$^3$-BO algorithm.}
\label{alg:scalable3bo}
\algorithmicinput dataset $\mathcal{D}_n = (\mathbf{x}, \mathbf{y})_{i=1}^n$, dimensionality $d > d_e$, max number of inducing points $m$, $f(\cdot)$ with scaled input in $[-\sqrt{d}, \sqrt{d}]^d$, bounds $[\underline{\mathbf{x}}, \overline{\mathbf{x}}]$

\begin{algorithmic}[1]

\State draw a random matrix $\mathbf{A} \in \mathbb{R}^{D\times d}: a_{ij} \overset{\text{i.i.d}}{\sim} \mathcal{N}(0,1)$  \Comment{$\mathbf{A}$ is a Gaussian random matrix}
\State set $\mathcal{Z} \subset \mathbb{R}^d = [-\sqrt{d}, + \sqrt{d}]^{d}$
\State $\mathcal{D}_0 \gets \varnothing$

\While{convergence criteria not met}
\While{computational budget is not available} \Comment{threshold the computational budget}
\State wait and check periodically if there is any update 
\EndWhile
\State update input, output, and status for all cases \Comment{if not complete then hallucinate}
\State update dataset $\mathcal{D}_{i}$
\State determine batch to fill \Comment{exploit/explore or purely explore}
\State locate next sampling point: $\mathbf{z}_{i+1} = \argmax_{\mathbf{z} \in \mathcal{Z}} a(\mathbf{z}) \in \mathbb{R}^d$
\State embed, normalize, scale, and translate: $\mathbf{x}^*_{i+1} \gets \underline{\mathbf{x}} + \frac{ \frac{1}{d} \mathbf{A} \mathbf{z}_{i+1} + \sqrt{d}^d}{2(\sqrt{d})^d}  \odot (\overline{\mathbf{x}} - \underline{\mathbf{x}}) $ \Comment{$\odot$ is Hadamard product}
\State project $\mathbf{x}_{i+1}$ to $\mathcal{X}$: $\mathbf{x}_{i+1} \gets p_{\mathcal{X}} (\mathbf{x}^*_{i+1})$ \Comment{$p_{\mathcal{X}}( \mathbf{z}) = \argmin_{\mathbf{x} \in \mathcal{X}} \lVert \mathbf{x} -  \mathbf{z}  \rVert_2$}
\State query $\mathcal{D}_{i+1} \gets \mathcal{D}_i \cup  \{ \mathbf{z}_{i+1}, f(\mathbf{x}_{i+1}) \} $ \Comment{decouple the query and the main optimizer}
\State hallucinate the sparse GP
\State sample inducing inputs $\mathbf{Z_u}$, where $| \mathbf{Z_u}| = \min\{|\mathbf{X}|, m\}$ \Comment{Latin hypercube sampling, $|\cdot|$ denotes cardinality}
\State update the sparse GP \Comment{fully independent condition sparse GP (Section \ref{subsubsec:FIC-SparseGP})}
\EndWhile

\end{algorithmic}
\end{algorithm*}

GPstuff \cite{vanhatalo2012bayesian,vanhatalo2013gpstuff} is used as the underlying sparse GP implementation, where ELBO is used to train sparse GP as described in Equation \ref{eq:ELBOSparseGP}. Even though we could also optimize the inducing variable $\mathbf{X_u}$ in the sparse GP formulation, in this paper, $\mathbf{X_u}$ is obtained using Latin hypercube sampling as a quasi-Monte Carlo approach to further relax the computational cost to train the sparse GP. 

Without loss of generality, we assume that $f(\cdot)$ is rescaled in $\mathcal{Z} = [-\sqrt{d}, \sqrt{d}]^d $, otherwise, it could be rescaled by $2 \sqrt{d}^{d} \frac{\mathbf{x} - \underline{\mathbf{x}}}{\overline{\mathbf{x}} - \underline{\mathbf{x}}} - \sqrt{d}^d$\footnote{The function $(b-a) \frac{{x} - \underline{{x}}}{\overline{{x}} - \underline{{x}}} + a$ is an affine map from $[\underline{x},\overline{x}]$ to $[a,b]$.}, where $\underline{\mathbf{x}}$ and $\overline{\mathbf{x}}$ are the lower- and upper-bound of $\mathbf{x}$, respectively. Algorithm \ref{alg:scalable3bo} summarizes the Scalable$^3$-BO approach proposed in this paper. We note that there is always a chance for the embedding contains or does not contains the optimum, thus multiple runs are required. Choosing the dimensionality $d$ also has an effect on this probability, as larger $d$ corresponds to higher chance of containing the optimum and slower to converge and vice versa \cite{letham2020re}.

\begin{figure}[!htbp]
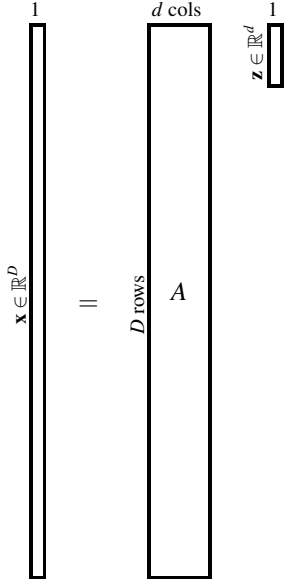

\[\renewcommand\matscale{0.25}
\matbox{70}{\LARGE $\mathbf{x} \in \mathbb{R}^{D}$\normalsize}{2}{\LARGE 1 \normalsize}{%
  \titlebox[20ex]{  }{  }} 
\quad = \quad
\matbox{70}{\LARGE $D$ rows \normalsize}{8}{\LARGE $d$ cols \normalsize}{\titlebox[10ex]{A}{ }} 
\quad 
\raiserows{31}{\matbox{8}{\LARGE $\mathbf{z} \in \mathbb{R}^{d}$ \normalsize}{2}{\LARGE 1 \normalsize}{\titlebox[10ex]{   }{  }}}
\]
\caption{A random embedding or a random projection $\mathbf{x} = \mathbf{A} \mathbf{z}$ is built as a corollary from the Johnson-Lindenstrauss lemma, where $\mathbf{A}$ is a random normal matrix.}
\label{fig:embeddingSchematic}
\end{figure}

Figure \ref{fig:embeddingSchematic} illustrates the concept of embedding from a low-dimensional space $\mathcal{Z}$ to a high-dimensional space $\mathcal{X}$. Since $\mathbf{x} = \mathbf{A} \mathbf{z}$, we could consider $x_i = \frac{1}{d} \sum_{j=1}^d a_{ij} z_j$, with $1 \leq i \leq D$. For  $d=1$, because $a_{ij} \overset{\text{i.i.d}}{\sim} \mathcal{N}(0,1)$ and $z_j \sim \mathcal{U}[-1,1]$, it is easy to show that $\mathbb{E}[x_i] = 0$ and $\mathbb{V}[x_i] = \frac{1}{3}$. Indeed, the probability density function of $x$ is $f_X(x) = \frac{1}{\sqrt{8\pi}} \Gamma\left( 0, \frac{x^2}{2} \right)$ for $x>0$ and $f_X(x) = f_X(-x)$ for $x<0$, where $\Gamma(\cdot)$ is the upper incomplete Gamma function. 
For $d>1$, if $z_j \sim \mathcal{U}[-\sqrt{d},\sqrt{d}]^d$, then $\mathbb{V}[x_i] = \frac{d^d}{3}$, but the probability density function of $x$ does not have any closed-form representation. Therefore, we perform a Monte Carlo simulation with $10^6$ samples to check what the probabilities of $x \in [-1,1]$ and $x \in [-\sqrt{d}, \sqrt{d}]^d$ are as a function of $d$, under the assumption that $z \sim \mathcal{U}[-\sqrt{d}, \sqrt{d}]^d$ is uniformly distributed. 

\begin{figure}[!htbp]
\includegraphics[width=0.5\textwidth, keepaspectratio]{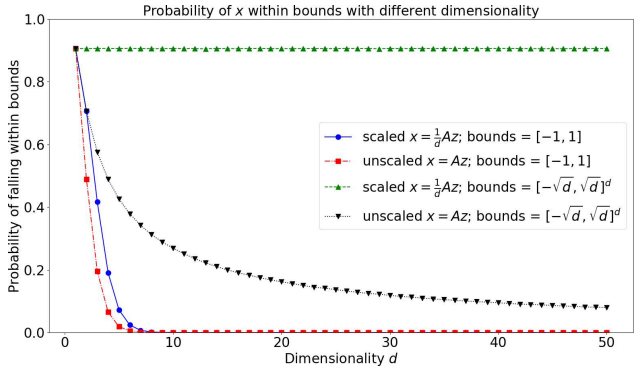}
\caption{Probability of $\mathbf{x} \in [-1,1]$ and $\mathbf{x} \in [-\sqrt{d}, \sqrt{d}]^d$ for $\mathbf{x} = \frac{1}{d} \mathbf{A} \mathbf{z}$ (scaled) and $\mathbf{x} = \mathbf{A} \mathbf{z}$ (unscaled) for $z \in \mathcal{Z} = [-\sqrt{d},\sqrt{d}]^d$ is uniformly distributed.}
\label{fig:cropped_bound-test-x}
\end{figure}

Figure \ref{fig:cropped_bound-test-x} shows the results of the Monte Carlo simulation, where nearly all the options are associated with zero probability. This means after the convex projection $p_\mathcal{X}(\cdot)$, nearly all the inputs will be located at boundaries, which is undesirable. Figure \ref{fig:cropped_bound-test-x} also justifies the bounds of $[-\sqrt{d},\sqrt{d}]^d$ for $\mathcal{X}$ before scaling and translation to the real bounds of $[\underline{\mathbf{x}}, \overline{\mathbf{x}}]$. Indeed, the probability of $\mathbf{x} = \frac{1}{d} \mathbf{A} \mathbf{z} \in [-\sqrt{d},\sqrt{d}]^d$ is approximately 90.6\%, which means approximately 9.4\% of the evaluation (under uniform prior assumption) will need the convex projection operator $p_\mathcal{X}(\cdot)$, which is not ideal but it is the best and stable choice compared to others, as shown Figure \ref{fig:cropped_bound-test-x}.

\section{Numerical benchmark}
\label{sec:numericalBenchmark}

We consider the second objective function in several problems of the ZDT test suite \cite{zitzler2000comparison} for multi-objective optimization problems, where the goal is to minimize $f_2(\mathbf{x})$, as well as a modified sphere function. We set the dimensionality $D=10,000$ to test out the capability of the proposed method. 

\subsection{Single-objective (modified) ZDT test suite}

To make the input space $\mathcal{X}$ centered around 0, we replace $\sum_{i=2}^D x_i$ with $\left( \sum_{i=2}^D x_i \right)^2$ to shift the domain from $[0,1]^D$ to $[-1,1]^D$. It is noted that in the ZDT test suite, there are only two active directions: $[1,0,\cdots,0]$ and $[0,1,\cdots,1]$, which correspond to $x_1$ and $\sum_{i=2}^D x_i$. The global optimal point for the unscaled ZDT test suite is $f(\mathbf{x^*}) = 0$, where the optimizer  is $\mathbf{x^*} = [1,0,\cdots,0] \in [0,1]^D$.

\subsubsection{ZDT1}
For a $D$-dimensional input $\mathbf{x} \in [0,1]^D$, a more general ZDT1 second-objective function can be described as \cite{zitzler2000comparison}
\begin{equation}
f_2(\mathbf{x}) = g \left( 1 - \sqrt{ \frac{x_1}{g} } \right),
\end{equation}
where $g = 1 + 9 \sum_{i=2}^{D} \frac{x_i}{D-1}$.

The modified ZDT1 function, which is defined on $[-1,1]^D$, is
\begin{equation}
f_2(\mathbf{x}) = g \left( 1 - \sqrt{ \frac{x_1^2}{g} } \right),
\label{eq:modZDT1}
\end{equation}
where $g = 1 + 9 \left( \sum_{i=2}^{D} \frac{x_i}{D-1} \right)^2$.

\subsubsection{ZDT2}
For a $D$-dimensional input $\mathbf{x} \in [0,1]^D$, the second-objective ZDT2 function can be described as \cite{zitzler2000comparison}
\begin{equation}
f_2(\mathbf{x}) = g \left[ 1 - \left( \frac{x_1}{g} \right)^2 \right],
\end{equation}
where $g = 1 + 9 \sum_{i=2}^{D} \frac{x_i}{D-1}$. 

The modified ZDT2 function, which is defined on $[-1,1]^D$, is
\begin{equation}
f_2(\mathbf{x}) = g \left[ 1 - \left( \frac{x_1}{g} \right)^2 \right],
\label{eq:modZDT2}
\end{equation}
where $g = 1 + \left( 9 \sum_{i=2}^{D} x_i \right)^2$. 

\subsubsection{ZDT3}
For a $D$-dimensional input $\mathbf{x} \in [0,1]^D$, the second-objective ZDT3 function can be described as \cite{zitzler2000comparison}
\begin{equation}
f_2(\mathbf{x}) = g \left[ 1 - \sqrt{\frac{x_1}{g}} - \left( \frac{x_1}{g} \sin(10\pi x_1) \right) \right],
\end{equation}
where $g = 1 + 9 \sum_{i=2}^{D} \frac{x_i}{D-1}$.

The modified ZDT3 function, which is defined on $\mathbf{x} \in [0,1]^D$, is
\begin{equation}
f_2(\mathbf{x}) = g \left[ 1 - \sqrt{\frac{x_1^2}{g}} - \left( \frac{x_1^2}{g} \sin(10\pi x_1^2) \right) \right],
\label{eq:modZDT3}
\end{equation}
where $g = 1 + 9 \left( \sum_{i=2}^{D} x_i \right)^2$


\subsection{Sphere function}

We consider a valley-shaped parabolic function
\begin{equation}
\label{eq:modifiedParabol}
f(\mathbf{x}) = \left(\sum_{i=1}^D x_i \right)^2
\end{equation}
on $[0,1]^{D}$, which has a 1d effective dimension of $\sum_{i=1}^D x_i $ along the direction $(1,\cdots,1)$. 
The test function \ref{eq:modifiedParabol} could be more generalized to 
\begin{equation}
f(\mathbf{x}) = \prod_{j=1}^{d_e} \left( \sum_{i=1}^D w_i^{(j)} x_i \right)^2
\end{equation}
with $d_e$ effective dimension in the direction of $e_j$, $1\leq j \leq d_e$, where $e_j = [w_1^{(j)}, \cdots, w_d^{(j)} ]$.

\section{Numerical results}

\subsection{Data scalability -- Effect of data size}

\begin{figure}[!htbp]
\includegraphics[width=0.455\textwidth, keepaspectratio]{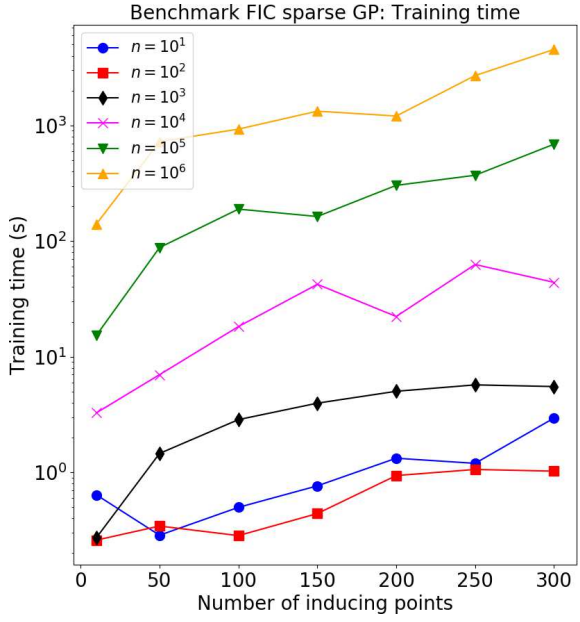}
\caption{Benchmark of training time.}
\label{fig:fic_benchmark_train}
\end{figure}

\begin{figure}[!htbp]
\includegraphics[width=0.455\textwidth, keepaspectratio]{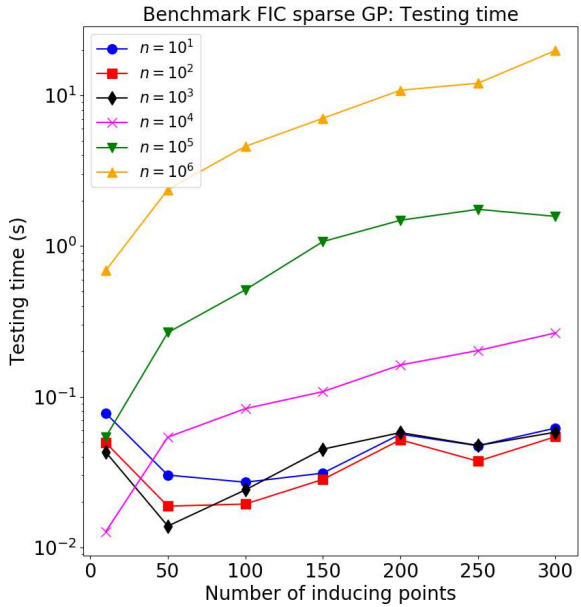}
\caption{Benchmark of testing time.}
\label{fig:fic_benchmark_test}
\end{figure}

\begin{figure}[!htbp]
\includegraphics[width=0.455\textwidth, keepaspectratio]{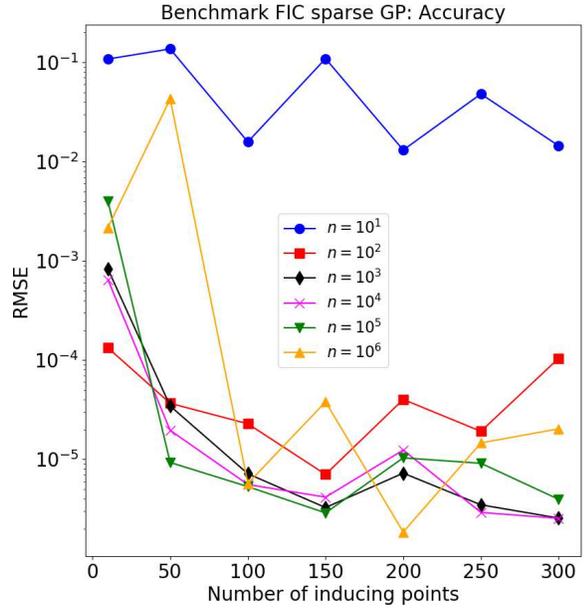}
\caption{Benchmark of accuracy.}
\label{fig:fic_benchmark_accuracy}
\end{figure}

Here we conduct a simple stress test for the FIC sparse GP framework used in this paper by considering the sphere function described in Equation \ref{eq:modifiedParabol}, where both the size of the dataset as well as the number of inducing points are varied. We compare both the training time, testing time, and testing accuracy to expose their trade-offs. The benchmark is conducted on Intel Xeon Platinum 8160 CPU @ 2.10GHz supported by RHEL 7.1 (Maipo) with 180 GB of memory. To be fair, the same test set of 1,000 points are held separately with respect the number of data points, where the training points and inducing points are chosen via Latin hypercube sampling. The dimensionality of the test is 3 on the domain of $[-1,1]^3$. The number of data points varies as $10^1, 10^2, \dots, 10^6$, with the maximum number of data points is set at 1M. The number of inducing points varies from $10, 50, 100, \dots, 300$, which is the rank of the low-rank matrix $\mathbf{K}_{\mathbf{u},\mathbf{u}}$. 
Figures \ref{fig:fic_benchmark_train}, \ref{fig:fic_benchmark_test}, and \ref{fig:fic_benchmark_accuracy} present the training time, testing time, and accuracy of the underlying sparse GP used in this Scalable$^3$-BO framework, respectively. At the extreme of $n=10^6$ and $m=300$, it takes approximately 49 minutes to fit the sparse GP. The root mean square error (RMSE) does not seem to decrease significantly for this particular 3D testing function and seem to have reached its limit at $10^{-5}$. However, more tests are needed before a conclusion can be drawn.

\subsection{Effect of dimensionality}

\begin{figure}[!htbp]
\includegraphics[width=0.455\textwidth, keepaspectratio]{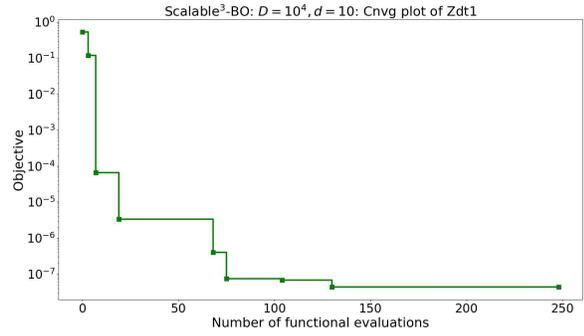}
\caption{ZDT1 (Equation \ref{eq:modZDT1}) convergence plot with $D=10,000$, $d=10$.}
\label{fig:cropped_Zdt1_cnvg}
\end{figure}

\begin{figure}[!htbp]
\includegraphics[width=0.455\textwidth, keepaspectratio]{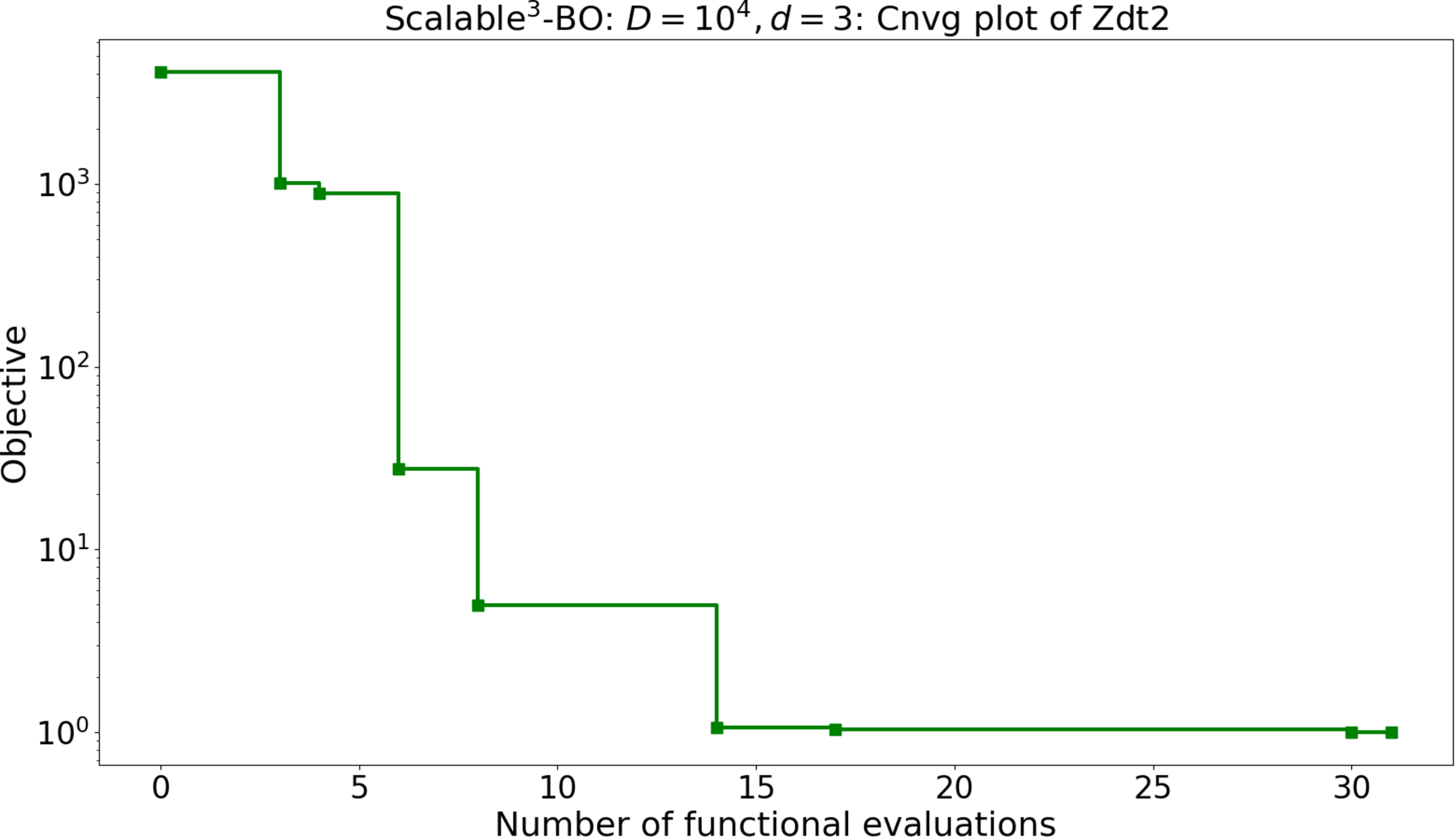}
\caption{ZDT2 (Equation \ref{eq:modZDT2}) convergence plot with $D=10,000$, $d=3$.}
\label{fig:cropped_Zdt2_cnvg}
\end{figure}

\begin{figure}[!htbp]
\includegraphics[width=0.455\textwidth, keepaspectratio]{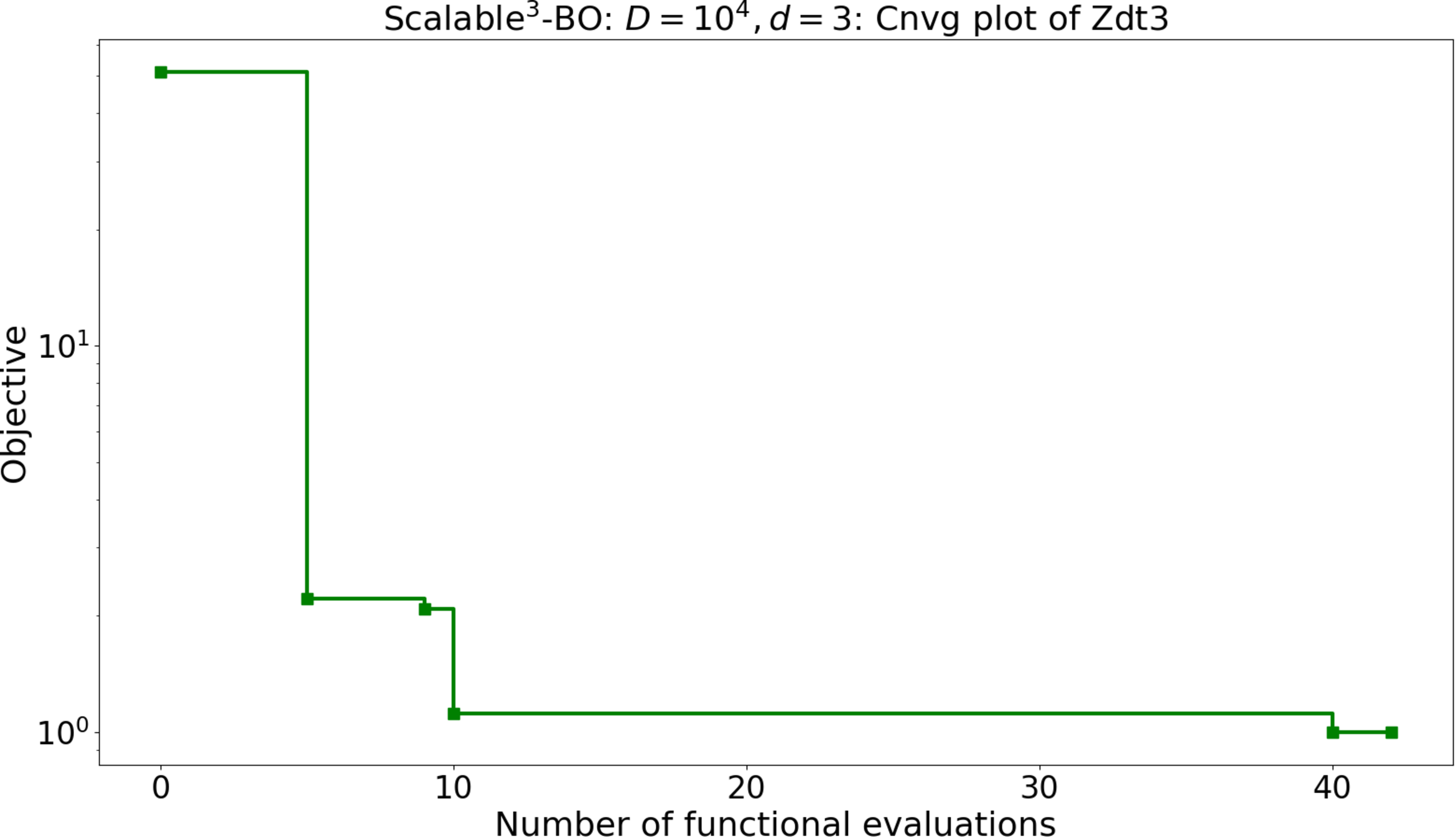}
\caption{ZDT3 (Equation \ref{eq:modZDT3}) convergence plot with $D=10,000$, $d=3$.}
\label{fig:cropped_Zdt3_cnvg}
\end{figure}

\begin{figure}[!htbp]
\includegraphics[width=0.455\textwidth, keepaspectratio]{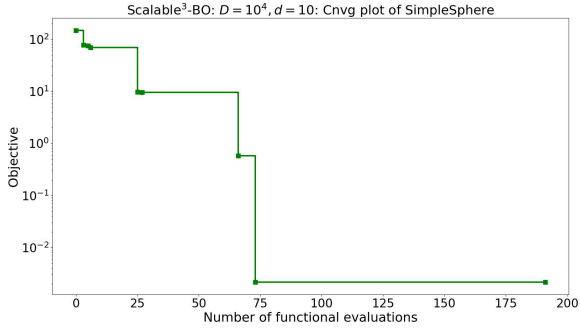}
\caption{Sphere (Equation \ref{eq:modifiedParabol}) convergence plot with $D=10,000$, $d=10$.}
\label{fig:cropped_SimpleSphere_cnvg}
\end{figure}

The proposed Scalable$^3$-BO method is benchmarked using the modified ZDT test suite, as well as the sphere function described in Section \ref{sec:numericalBenchmark}. 
Here, the Mat\'{e}rn kernel $\nu = 3/2$, $k(\mathbf{x}, \mathbf{x}') = \theta_0^2 \exp(-\sqrt{3}r)(1+\sqrt{3}r)$, is used. 
Figures \ref{fig:cropped_Zdt1_cnvg}, \ref{fig:cropped_Zdt2_cnvg}, \ref{fig:cropped_Zdt3_cnvg}, and \ref{fig:cropped_SimpleSphere_cnvg} describe the convergence plot of the modified ZDT1, ZDT2, ZDT3, and sphere benchmarking functions, respectively. 
For the ZDT test suite, the dominant factor is $g$, and while the global optimum could be close to 0, we are more interested in the convergence rate and how well the proposed framework performs in pessimistically high-dimensional problems. In that sense, a fast convergence to $10^0$ on the scale of the objective function in a short time can be considered as a fair achievement. 

\subsection{Effect of batch size -- Computing scalability}

In this section, we examine the scalability and performance of the asynchronous parallel feature in Scalable$^3$-BO with a standard 4D Hartmann function. The objective function is described as 
\begin{equation}
f(\mathbf{x}) = \frac{1}{0.839} \left[ 1.1 - \sum_{i=1}^4 \alpha_i \exp\left( - \sum_{j=4}^3 A_{ij} (x_j - P_{ij})^2 \right) \right],
\end{equation}
where $\mathbf{A} = \begin{pmatrix} 10 & 3 & 17 & 3.50 & 1.7 & 8 \\ 0.05 & 10 & 17 & 0.1 & 8 & 14 \\ 3 & 3.5 & 1.7 & 10 & 17 & 8 \\ 17 & 8 & 0.05 & 10 & 0.1 & 14 \end{pmatrix}$, $\mathbf{P} = 10^{-4} \begin{pmatrix} 1312 & 1696 & 5569 & 124 & 8283 & 5886 \\ 2329 & 4135 & 8307 & 3736 & 1004 & 9991 \\ 2348 & 1451 & 3522 & 2883 & 3047 & 6650 \\ 4047 & 8828 & 8732 & 5743 & 1091 & 381 \end{pmatrix}$, and $\alpha = (1.0, 1.2, 3.0, 3.2)^T$ on the domain of $[0,1]^4$. To mimic the computational expensive cost of real-world engineering simulations, we impose a uniformly distributed random computational cost, i.e. $t\sim \mathcal{U}[\underline{t}, \overline{t}]$. In this example, we set $\underline{t} = 30$s, $\overline{t} = 900$s, and study the convergence behavior as a function of workers. We compare the performance in terms of physical wall-clock waiting time. To be fair, all runs start with the exactly same number (i.e. two) and locations of initial sampling points. 

\begin{figure}[!htbp]
\includegraphics[width=0.455\textwidth, keepaspectratio]{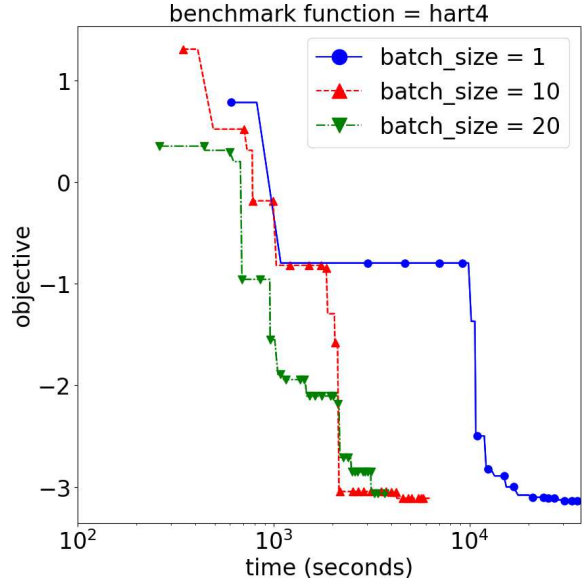}
\caption{Effect of batch size on the wall-clock time.}
\label{fig:cropped_batchsizeEffect}
\end{figure}

\begin{figure}[!htbp]
\includegraphics[width=0.455\textwidth, keepaspectratio]{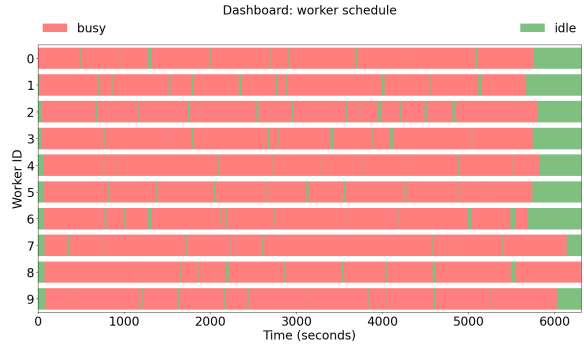}
\caption{Worker schedule for asynchronous parallel BO with batch size of 10.}
\label{fig:cropped_hart4-WorkerSchedule}
\end{figure}

Figure \ref{fig:cropped_batchsizeEffect} shows the superior numerical performance as larger batch size yield a better optimization performance in  term of time, statistically speaking. The classical sequential BO (with batch size of 1) simply takes much longer compared to its parallel variant for a slightly expensive application. Figure \ref{fig:cropped_hart4-WorkerSchedule} shows a typical schedule for asynchronous parallel BO with batch size of 10, where workers is nearly continuously assigned jobs, except for the small downtime of fitting GP and searching for the next sampling point. Therefore, the idle time for workers is minimized, further enhancing the effectiveness and efficiency of BO algorithms.

\section{Discussion}

To some extent, Scalable$^3$-BO fully enables computationally expensive applications, if there is sufficient computing resources. With the usage of the sparse GP, the number of data points is no longer a primary issue with expensive applications. However, one should be careful when encountering high-dimensional applications, as the employment of the sparse GP alone does not guarantee optimization efficiency on high-dimensional problems.

The Scalable$^3$-BO framework can be considered as the generalized parallel extension of the classical BO. To recover the classical and sequential BO algorithm, one can
\begin{itemize}
\item - set $\mathbf{A} = \mathbf{I}_{D \times D}$ to not use the random embedding feature,
\item - set number of worker to 1,
\item - assign the inducing points to be exactly the same with the training inputs. 
\end{itemize}
Also, it is non-trivial to consider the impact for each of these cases. Some theoretical results provided by Desautels et al.~\cite{desautels2014parallelizing}, Contal et al. \cite{contal2013parallel} and Gupta et al. \cite{gupta2018exploiting} gave more insight to the batch cumulative regrets at the order of $\mathcal{O}\left( \sqrt{ \frac{T\log T}{B} } \right)$ for GP-UCB-PE and of $\mathcal{O}\left( \sqrt{ \frac{T\log (TB)}{B} } \right)$ for GP-BUCB, where $T$ is the number of iterations and $B$ is the batch size. Therefore, the cumulative regret is sublinear, and because GP-BUCB is a no-regret algorithm, Scalable$^3$-BO is also a no-regret algorithm as well since it is derived from GP-BUCB and GP-UCB-PE. 

Comparing to the local GP approaches, so called the subset of data approach \cite{quinonero2005unifying,quinonero2007approximation}, the low-rank sparse GP approximation is more competitive in terms of theoretical formulation and predictive accuracy, with computational complexity of $\mathcal{O}(nm^2)$. The loss of accuracy in local GP approaches is mainly attributed to the vague correlation between data in different clusters or subsets of data. Nevertheless, both approaches remain interesting and beg for further research.

BO is a flexible optimization framework that naturally lends itself to multi-fidelity \cite{tran2020smfbo2cogp,tran2019sbfbo2cogp}, multi-objective \cite{tran2020srmobo3gp,tran2020srmobo3gp2} optimization problems with applications to computational solid mechanics \cite{tran2020multi,tran2020solving,tran2020an} and fluid mechanics \cite{tran2018weargp}. It is also worthy to mention that the asynchronous parallel feature for HPC is currently supported in DAKOTA \cite{dalbey2020dakota}. 

A closely related method for linearly reducing dimensionality is active subspace, which carefully investigates the spectral decomposition of $\mathbb{E}[ \nabla f(\mathbf{x}) \nabla f(\mathbf{x})^{\top} ]$ by singular value decomposition. Typically, how fast the eigenspectrum of $\mathbb{E}[ \nabla f(\mathbf{x}) \nabla f(\mathbf{x})^{\top} ]$ decays can be revealed by the singular value decomposition, where eigenvalues are arranged in descending order. Theoretical guarantees allow truncation of eigenvalues and subsequently dimensionality reduction within a controlled tolerance as usually done in active subspace \cite{constantine2014active,constantine2015active}. However, the goal of active subspace is to accurately approximate a function using a reduced basis, whereas the goal of random embedding in BO is to optimize under an unknown subspace. While these two concepts can be cross-cutting, random embedding is arguably more efficient because it does not require the BO method to discover the active subspace on-the-fly. Additive GP for high-dimensional problems offer a similar flavor, but this method directly targets the kernel $k(\mathbf{x},\mathbf{x'})$ instead of the inputs $\mathbf{x}$ itself.

\section{Conclusion}

In this work, we introduce Scalable$^3$-BO -- a robust and scalable BO framework implemented with the sparse GP, random embedding, and asynchronous parallelization -- to tackle data scalability issue, high-dimensional problems on HPC platforms for computationally expensive application. We demonstrate the scalability of the proposed framework up to 1M data points, 10,000D problems with small effective dimensionality, and 20 workers in HPC environment.

\section*{Acknowledgment}
This work was supported by the US Department of Energy, Office of Advanced Scientific Computing Research, Field Work Proposal 20-023231. Sandia National Laboratories is a multimission laboratory managed and operated by National Technology and Engineering Solutions of Sandia, LLC., a wholly owned subsidiary of Honeywell International, Inc., for the U.S. Department of Energy's National Nuclear Security Administration under contract DE-NA-0003525. The views expressed in the article do not necessarily represent the views of the U.S. Department of Energy or the United States Government.

%
%

\bibliographystyle{asmems4}
\bibliography{lib}   




\end{document}